\documentclass[12pt]{article}
\usepackage{graphicx}

\newcommand{\epem}{\ensuremath{\mathrm{e^+e^-}}}
\newcommand{\roots}  {\ensuremath{\sqrt{s}}}

\newcommand{\gevcc}{\mbox{GeV$/c^2$}}
\newcommand{\evcc}{\mbox{eV$/c^2$}}

\newcommand{\tev}{\mbox {TeV}}
\newcommand{\neut}{\ensuremath{\tilde{\chi}^0_1}}
\newcommand{\grav}{\ensuremath{\mathrm{\tilde{G}}}}

\newsavebox{\sboxpubnumber}
\newsavebox{\sboxpubdate}
\newcommand{\pubdate}[1]{\begin{lrbox}{\sboxpubdate}{#1}\end{lrbox}}
\newcommand{\pubnumber}[1]{\begin{lrbox}{\sboxpubnumber}
    {\begin{tabular}{l} #1 \\
        \usebox{\sboxpubdate}
      \end{tabular}}
  \end{lrbox}
  \pubblock}
\newcommand{\Title}[1]{\begin{center} {\Large #1 } \end{center}}
\newcommand{\Author}[1]{\begin{center}{ \sc #1} \end{center}}
\newcommand{\Address}[1]{\begin{center}{ \it #1} \end{center}}

\newcommand{\pubblock}{\rightline{
    \usebox{\sboxpubnumber}}}
\newenvironment{Abstract}{\begin{quotation}  }{\end{quotation}}
\newenvironment{Presented}{\begin{quotation} \begin{center}
      PRESENTED AT\end{center}\bigskip
    \begin{center}\begin{large}}{\end{large}\end{center}
  \end{quotation}}
\newcommand{\Acknowledgements}{\bigskip  \bigskip \begin{center} \begin{large}
      \bf ACKNOWLEDGEMENTS \end{large}\end{center}}

\begin{document}

\begin{titlepage}
\pubdate{November 27, 2001}                    
\pubnumber{hep-ex/0111085} 

\vfill
\Title{Searches for extra dimensions, gauge mediated SUSY and exotics at
  LEP}
\vfill
\Author{David Hutchcroft}
\Address{CERN \\
  CH-1211 Gen\`eve 23 \\
  Switzerland}
\vfill
\begin{Abstract}
  
  The results of searches for several type of physics beyond the
  Standard Model using data from the four LEP experiments are presented.
  In the absence of any excess signal events seen in the data limits are
  placed on the existence of extra-dimensions, gauge mediated
  supersymmetry and some exotic states.

\end{Abstract}
\vfill
\begin{Presented}
    COSMO-01 \\
    Rovaniemi, Finland, \\
    August 29 -- September 4, 2001
\end{Presented}
\vfill
\end{titlepage}
\def\thefootnote{\fnsymbol{footnote}}
\setcounter{footnote}{0}

\section{Introduction}

LEP was the large electron positron collider at CERN. This accelerator
ran from 1989 to 2000 producing \epem\ collisions at $\roots =91-208$
GeV in the four detectors around the ring. Each of the detectors was
used to measure the parameters of the Z boson precisely and that of
other standard model parameters. The large number of Z decays and other
measurements allow accurate tests of the standard model and constrain new
physics processes that would cause deviations from the standard model
predictions. 

Searches for new particles were performed by direct searches and
indirectly using the precision measurements. A brief overview of a few
of these searches is presented here.

\section{The LEP experiments}

The 4 LEP experiments, ALEPH, DELPHI, L3 and OPAL, use different
technologies based around the same basic detector design, as described
in \cite{ALEPH_Detector, DELPHI_Detector, L3_Detector, OPAL_Detector}.
They all rely on precise vertexing with silicon detectors around the
beam pipe, tracking charged particles within a solenoidal magnetic field
and measuring energy deposited in the electro-magnetic and hadronic
calorimeters and around the outside of the calorimeters dedicated muon
detectors.

\section{Extra Dimensions}

One possible solution to the problem of quantising gravity is to add
additional dimensions to the 3+1 space-time dimensions assumed in the
standard model
\cite{quantum_gravity1,quantum_gravity2,quantum_gravity3}. These extra
dimensions are finite in size and so will have quantised energy states
within them.  Gravitons, the particle that mediates the gravitational
force, propagates in these compactified extra dimensions. This can solve
the hierarchy problem, which is the very large scale difference between
the Plank scale ($M_{Pl} \sim {\cal O}(10^{18-19})$ GeV) and the
electro-weak scale ($M_{EW} \sim {\cal O}(10^{2-3})$ GeV). In a space
with $D=\delta + 4$ dimensions the Planck mass, $M_D$, will be modified
by the extra-dimensions. If the extra dimensions are of size $R$
\begin{equation}
G^{-1}_N = 8\pi R^\delta M^{2+\delta}_D,
\end{equation}
where $G_N$ is the Newtonian gravitational constant.

\begin{figure}[ht]
  \begin{center}
    \includegraphics[width=0.7\textwidth]{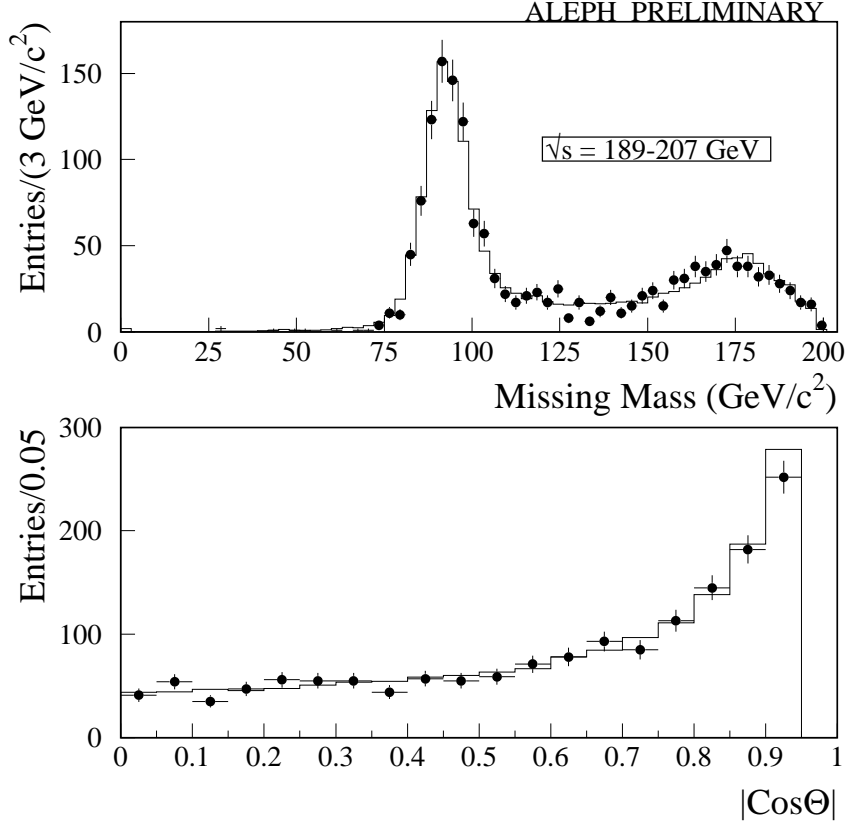}
    \caption{ALEPH data (points) and SM expectation (line) for the
      missing mass and $|\cos{\theta}|$ distributions.}
    \label{fig:ALEPH_photons}
  \end{center}
\end{figure}

In an \epem\ collider at electro-weak scales a signature of this process
is the production a photon with a graviton, $\epem\rightarrow G\gamma$.
The graviton can not be detected and the visible decay is a single
photon and missing energy and momentum. In the ALEPH data there was no
excess seen in the single photon distribution over the expected SM
backgrounds, see Figure \ref{fig:ALEPH_photons}. The largest possible
size extra dimension is a function of the number of assumed extra
dimensions. Assuming 2 extra dimensions, which gives the least stringent
size limit, they must be smaller than 0.29\,mm at $95\%$ confidence
which is equivalent to a lower bound of 1.28 TeV on $M_D$, see
reference \cite{ALEPH_photons} for the complete analysis.

Another signature is the modification of the 2 fermion production
processes through a virtual graviton, $\epem\rightarrow G^* \rightarrow
\overline{f}f$. Comparisons of the standard model distributions and
those with corrections for the extra dimensions can be made with the
data. The largest effect will be seen in Bhabha scattering:
\begin{equation}
  {d\sigma \over d\cos\theta} = A(s,t) + {\lambda \over M^4_s} B(s,t)
  + {\lambda^2 \over M^8_s} C(s,t),
\end{equation}
where $s$ is the square of the centre-of-mass energy, $t=-{1 \over
  2}s(1-\cos\theta)$ with $\cos\theta$ the electron scattering angle.
$A$ is the SM term, $B$ is the interference term and $C$ is the graviton
exchange term. $M_s$ is a mass scale related to $M_D$ and $\lambda$ is a
parameter that depends on the theory. Using OPAL data and fitting for
ee, $\mu\mu$, $\tau\tau$, $\gamma\gamma$ and ZZ distributions a
measurement of $\lambda/M^4_s = 0.31 \pm 0.39 \tev^{-4}$ was made, see
reference \cite{OPAL_gravitons} for the complete analysis.

\section{Gauge mediated SUSY breaking}
Minimal supersymmetry models (MSSM) assume that supersymmetry is broken
in the gravitational sector at high scale ($>>1$ TeV). If supersymmetry
is broken by gauge forces (GMSB) then the gravitino will be a light
particle, $M_{\tilde{G}} < 1\, \evcc$, and the lightest supersymmetric
particle. Decays of the next to lightest supersymmetric particle (NLSP)
to the gravitino are possible and may have a significant life time
($c\tau > 10$ cm). NLSP signatures in three lifetime ranges are shown in
Table \ref{tab:gmsb_topologies}.

\begin{table}[htbp]
  \begin{center}
    \begin{tabular}[h]{llcc}
      \hline
      \multicolumn{2}{c}{NLSP Lifetime} & sleptons & neutralinos \\
      \hline
      \hline
      Short & ($c\tau < 10$ cm) & Acoplanar & Acoplanar \\
                             &   & leptons & photons \\
      Medium &(5 cm $< c\tau < 10$ cm) & Track kinks & Single non-pointing \\
                              &         &             &  photons \\
      Long & ($c\tau > 2$ m) & Heavy stable  &        Invisible\\
                           &  & charged particles &   \\
                           &  & ($dE/dx$)&\\
      \hline
    \end{tabular}
    \caption{GMSB signatures with a slepton or neutralino NLSP}
    \label{tab:gmsb_topologies}
  \end{center}
\end{table}

\subsection{Two photons and missing energy}
\begin{figure}[htb]
  \begin{center}
    \includegraphics[width=\textwidth]{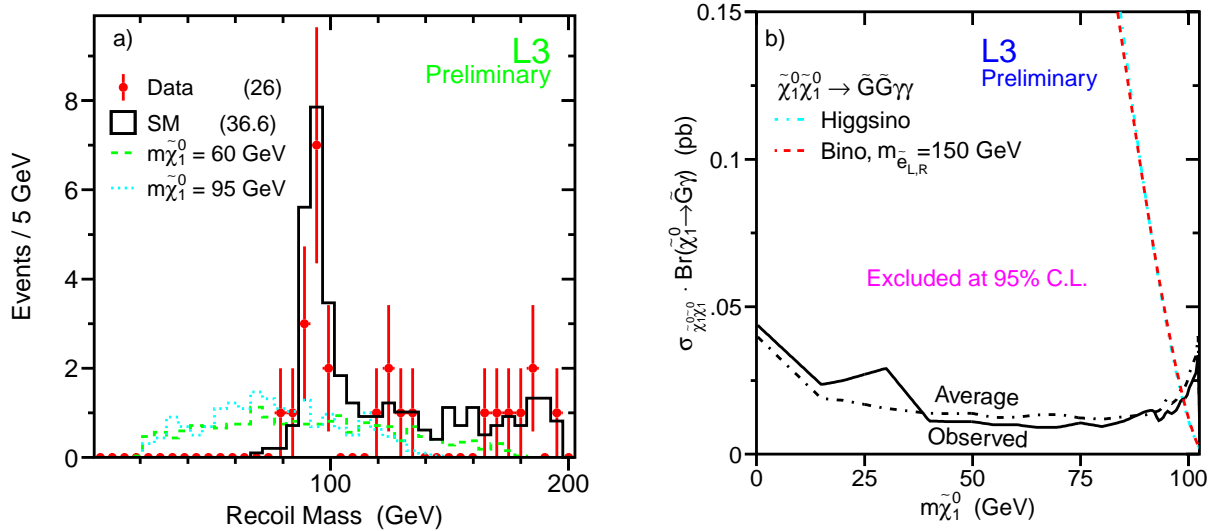}
    \caption{Plot a: L3 data and SM background predictions for the recoil mass
      from two photon events, two different mass neutralino predictions
      are also shown.
      Plot b: the cross-section $\times$ branching ratio exclusions for
      $\neut\neut\rightarrow\grav\grav\gamma\gamma$ and the
      cross-sections for two extreme case neutralino mixings.}
    \label{fig:L3_photons}
  \end{center}
\end{figure}

The search for excess two acoplanar photons and missing energy
constrains the neutralino pair production cross-section in the case of
rapid decay to gravitinos and photons. Using the LEP combined data a
limit on the production cross-section for neutralino pairs can be set.
Figure \ref{fig:L3_photons} for an example data and background
comparison and corresponding limits, see reference
\cite{L3_susy,LEP:two_photon} for a discussion of the complete analyses.

\subsection{Heavy stable charged particles}
\begin{figure}[htb]
  \begin{center}
    \includegraphics[width=0.75\textwidth]{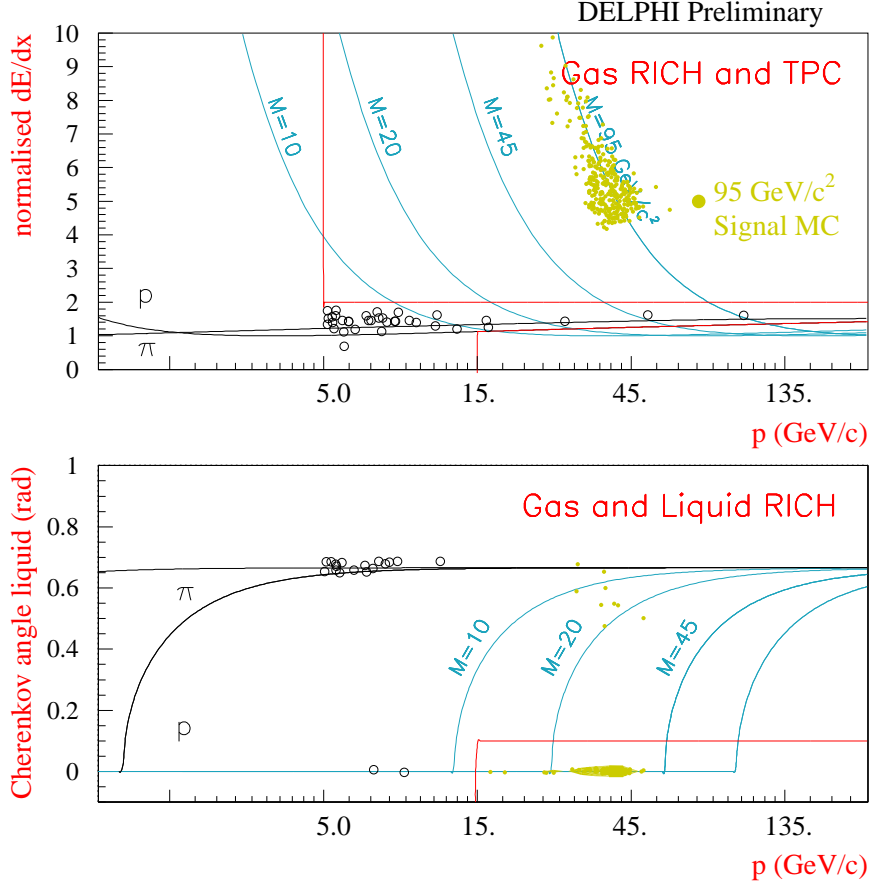}
    \caption{Normalised $dE/dx$ and \v Cerenkov angle distributions for
      DELPHI data and predictions for several signal heavy lepton masses.}
    \label{fig:DELPHI_stable_charged}
  \end{center}
\end{figure}

Heavy particles decaying with a distance of greater than 10\,m will be
seen as tracks with very high specific ionisation along their path and
will pass through the muon chamber when leaving the detectors. In DELPHI
the additional information given by the \v Cerenkov detectors is an
additional discriminant. These particles can be the signature of several
types of new physics including stable lepton species, see section
\ref{sec:heavy_lepton}, NLSP states in GMSB or MSSM states with small
R-parity violating couplings. Figure \ref{fig:DELPHI_stable_charged}
shows the DELPHI data and Monte Carlo prediction for several stable
charged particle masses for $dE/dx$ and \v Cerenkov angles, for a full
analysis see \cite{DELPHI_GMSB}.  This data allows the setting of a
cross-section limit as a function of the mass of the proposed particle,
for stable smuons or staus the limit corresponds to 97.4(97.1)\,GeV for
left(right) handed sleptons \cite{DELPHI_GMSB}. For selections
production the t-channel processes depend on the masses of the
neutralinos and so a general limit can not be set.

\subsection{Track kink searches}
If a particle has a decay length of the order of the size detector it
will be characterised by decays in the bulk of the detector,
characterised by a kink in the track. Backgrounds to these topologies
are usually caused by nuclear interactions, where a track elastically
collides with a nucleus, and cosmic rays in which particles,
predominantly muons, arriving out of time with a beam crossing are
reconstructed in the wrong positions.  These searched when combined with
the long and short lifetime searches can completely cover the parameter
space with a slepton NLSP. The limits set by the OPAL data is that for
all lifetimes cross-sections of larger than 0.1\,pb are excluded for
each of the three slepton flavours, see Reference \cite{OPAL_GMSB} for a
complete discussion of the analysis.

\section{Exotic Leptons}
\label{sec:heavy_lepton}
Exotic leptons can be of several types depending on the extension to the
standard model assumed. They can be classified according to their
$SU(2)\times U(1)$ quantum numbers \cite{EW_extraleptons}. Three
possible cases are
\begin{list}{}{}
\item[Sequential leptons \cite{seq_lepton}:] a fourth family of standard
  model leptons is assumed,
\item[Vector leptons \cite{vector_lepton}:] which are vector particles
  and occur in left and right handed Iso-spin doublets,
\item[Mirror leptons \cite{mirror_lepton}:] which have opposite chiral
  properties to the SM leptons.
\end{list}
Using the data collected by the L3 experiment with the assumptions that
the charged current mode dominates, which is valid for the kinematically
accessible regions, and a short ($<$ 1\,cm) lifetime or long lifetime the
limits set are listed in Table \ref{tab:l3_exotic_leptons}.
Reference \cite{L3_heavy_leptons} has a complete discussion of the
analysis and limits set.

\begin{table}[htb]
\begin{center}
\begin{tabular}{c c c c}
\hline
 & Sequential & Vector & Mirror \\
\hline
\hline
 $L^0\rightarrow\tau \mathrm{W}$ (Dirac) & 90.3 & 99.3 & 90.3 \\
 $L^0\rightarrow\tau \mathrm{W}$ (Majorama) & 80.5 & -- & 80.5 \\ 
 $L^\pm\rightarrow\nu \mathrm{W}$ & 100.8 &101.2 &100.5 \\
 $L^\pm\rightarrow L^0 \mathrm{W}$ & 101.9 &102.1 &101.9 \\
 Stable $L^\pm$ & 102.6 &102.6 &102.6 \\
\hline
\end{tabular}
\end{center}
\caption{Lower limits in \gevcc\ on the masses of exotic lepton species
  set using the L3 data.}
\label{tab:l3_exotic_leptons}
\end{table}

\section{Technicolor}
Technicolor \cite{technicolor} is an alternative to the Higgs mechanism
for electro-weak symmetry breaking. A new charge which is similar to the
colour charge in QCD is proposed. This would be carried by a set of new
particles called technifermions. It is the breaking of the chiral
symmetry associated to these which gives masses to the Standard Model
bosons. The simplest models are ruled out by precision electro-weak
measurements \cite{technicolor_exclusion} and so more complex models
that avoid these limits are proposed. One way is to require that the
coupling parameter changes more slowly with scale that that of QCD,
rather than a running parameter it is described as a walking one. The
most basic technicolor model used to set exclusions in LEP searches is
the straw man model \cite{techincolor_strawman}, in this model the
lightest technicolor vector mesons could be seen at LEP. 

\begin{figure}[htb]
  \begin{center}
    \includegraphics[width=\textwidth]{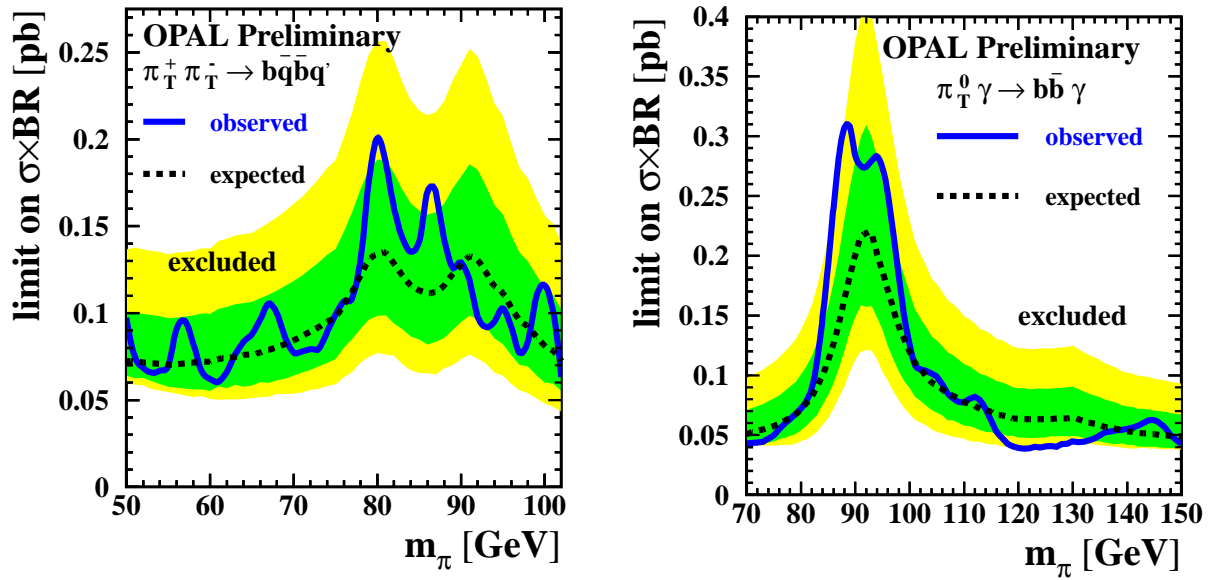}
    \caption{Limits set with the OPAL data on two topologies for
      technicolor production. The blue curve is the upper limit on
      technicolor production, the dashed curve is the expected limit and
      the green and yellow bands are the 1$\sigma$ and 2$\sigma$ curves
      around this.}
    \label{fig:OPAL_techin_cs}
  \end{center}
\end{figure}

The two channels used to search for technicolor decays are
\begin{equation}
  \epem \rightarrow \rho^0_T /\omega^0_T \rightarrow \pi^+_T\pi^-_T 
  \rightarrow b\overline{q}\overline{b}q'
\end{equation}
\begin{equation}
  \epem \rightarrow \rho^0_T /\omega^0_T \rightarrow \pi^0_T\gamma 
  \rightarrow b\overline{b}\gamma
\end{equation}
which are resonant production processes. The cross-section is much
greater if $M(\rho^0_T/\omega_T^0)\sim\sqrt{s}$, although there is some
sensitivity at other centre-of-mass energies through radiative events
and by producing virtual techinparticles. Cross-section limits on the
processes listed above are shown in Figure \ref{fig:OPAL_techin_cs}
\cite{OPAL:Techincolor}.  Limits are set using both OPAL and DELPHI data
on possible parameters of the see references
\cite{OPAL:Techincolor,DELPHI:Technicolor} for a complete discussion of
the analysis and limits. 

\section{Conclusions}

The data taken from 1989--2000 by the four LEP experiments have been
analysised and many searches for physics beyond the standard model have
been performed. So far there has been no evidence for any such process.
This is only a partial survey of the large number of direct and indirect
searches performed by the LEP collaborations.

\Acknowledgements
We wish to express our gratitude to the CERN accelerator divisions for
the excellent performance of the LEP machine. We acknowledge with
appreciation the efforts of the engineers, technicians and support
staff who have participated in the construction and maintenance of these
experiments.


\begin{thebibliography}{99}


\bibitem{ALEPH_Detector}
D.~Decamp {\it et al.}  [ALEPH Collaboration],
``Aleph: A Detector For Electron - Positron Annihilations At LEP,''
Nucl.\ Instrum.\ Meth.\ A {\bf 294} (1990) 121
[Erratum-ibid.\ A {\bf 303} (1990) 393].


\bibitem{DELPHI_Detector}
P.~Aarnio {\it et al.}  [DELPHI Collaboration],
``The DELPHI detector at LEP,''
Nucl.\ Instrum.\ Meth.\ A {\bf 303} (1991) 233.


\bibitem{L3_Detector}
  [L3 Collaboration],
``The Construction Of The L3 Experiment,''
Nucl.\ Instrum.\ Meth.\ A {\bf 289} (1990) 35.

\bibitem{OPAL_Detector}
K.~Ahmet {\it et al.}  [OPAL Collaboration],
``The OPAL detector at LEP,''
Nucl.\ Instrum.\ Meth.\ A {\bf 305} (1991) 275.

\bibitem{quantum_gravity1}
N.~Arkani-Hamed, S.~Dimopoulos and G.~R.~Dvali,
Phys.\ Lett.\ B {\bf 429} (1998) 263
[arXiv:hep-ph/9803315].

\bibitem{quantum_gravity2}
I.~Antoniadis, N.~Arkani-Hamed, S.~Dimopoulos and G.~R.~Dvali,
Phys.\ Lett.\ B {\bf 436} (1998) 257
[arXiv:hep-ph/9804398].

\bibitem{quantum_gravity3}
N.~Arkani-Hamed, S.~Dimopoulos and G.~R.~Dvali,
Phys.\ Rev.\ D {\bf 59} (1999) 086004
[arXiv:hep-ph/9807344].

\bibitem{ALEPH_photons}
ALEPH Collaboration,
``Single- and multi-photon production and a search for slepton pair production in GMSB topologies in e+e- collisions at sqrt(s) up to 208 GeV,''
ALEPH 2001-010, CONF 2001-007

\bibitem{OPAL_gravitons}
OPAL Collaboration,
``Limits on Low Scale Quantum Gravity in Extra Spatial Dimensions from
measurements of $e^+e^- \rightarrow e^+e^-$ at LEP2,''
OPAL Physics Note PN471

\bibitem{L3_susy}
L3 Collaboration,
``Search for Supersymmetry in $e^+e^-$ collisions at $\sqrt{s}=202-208$
GeV,''
L3 Note 2707

\bibitem{LEP:two_photon}
LEPSUSYWG, ALEPH, DELPHI, L3 and OPAL experiments, 
``Acoplanar two-photon events,''
LEPSUSYWG/01-04.1 

\bibitem{DELPHI_GMSB}
DELPHI Collaboration,
``Search for supersymmetric particles in light gravitino scenarios,''
DELPHI 2001-075, CONF 503

\bibitem{OPAL_GMSB}
OPAL Collaboration,
``Searches for Intermediate Lifetime Signatures in GMSB Models with a
Slepton NLSP in $e^+e^-$ collisions at $\sqrt{s}=189-208$,''
OPAL Physics Note PN478

\bibitem{EW_extraleptons}
For a review see:
\newline
D. London, in ``Precision Tests of the Standard Model,'' ed
P. Langacker, World Scientific, Singapore (1995); 
\newline
A. Djouadi, J. Ng and T.G. Rizzo, in ``Electroweak Symmetry Breaking and
New Physics at the TeV Scale,'' eds T. Barklow et al. World Scientific,
Singapore (1997)

\bibitem{seq_lepton}
M.~Maltoni, V.~A.~Novikov, L.~B.~Okun, A.~N.~Rozanov and M.~I.~Vysotsky,
Phys.\ Lett.\ B {\bf 476} (2000) 107
[arXiv:hep-ph/9911535].

\bibitem{vector_lepton}
For a review see: J. Hewett and T.G. Rizzo, Phys. Rep. 183, (1989) 193

\bibitem{mirror_lepton}
J.~Maalampi and M.~Roos,
Phys.\ Rept.\  {\bf 186} (1990) 53.

\bibitem{L3_heavy_leptons}
L3 Collaboration,
``Search for Heavy Neutral and Charged Leptons in $e^+e^-$ Annihilation
at LEP,''
Submitted to Phys. Lett. B,
CERN-EP/2001-046

\bibitem{technicolor}
S. Wienberg, Phys. Rev. {\bf D13} (1976) 974;
\newline
S. Wienberg, Phys. Rev. {\bf D19} (1979) 1277;
\newline
L. Susskind, Phys. Rev. {\bf D20} (1979) 2619

\bibitem{technicolor_exclusion}
M.E. Peskin, T. Takeuchi, Phys. Rev. {\bf D46} (1992) 381;
\newline
J. Erler, P. Langacker, Eur. Phys. J. {\bf C 15} (2000) 95

\bibitem{techincolor_strawman}
K.~D.~Lane,
Phys.\ Rev.\ D {\bf 60} (1999) 075007
[arXiv:hep-ph/9903369].

\bibitem{OPAL:Techincolor}
OPAL Collaboration,
``Searches for Technicolor with the OPAL Detector in $e^+e^-$ Collisions
at the Highest LEP Energies,''
OPAL Physics Note PN485

\bibitem{DELPHI:Technicolor}
DELPHI Collaboration,
``Search for Technicolor with DELPHI,''
DELPHI 2001-086 CONF 514

\end{thebibliography}
\end{document}